\documentclass[]{aastex}

\slugcomment{}

\shorttitle{Variation of Inner Dust Torus in NGC 4151}
\shortauthors{Koshida et al.}

\begin{document}

\title{Variation of Inner Radius of Dust Torus in NGC4151}

\author{Shintaro Koshida\altaffilmark{1,2}, Yuzuru Yoshii
\altaffilmark{3,4}, 
Yukiyasu Kobayashi\altaffilmark{2}, Takeo Minezaki\altaffilmark{3},
Yu Sakata\altaffilmark{1,3}, Shota Sugawara\altaffilmark{1,3},
Keigo Enya\altaffilmark{5}, Masahiro Suganuma\altaffilmark{2}, Hiroyuki 
Tomita\altaffilmark{3}, Tsutomu Aoki\altaffilmark{6}, and Bruce A. 
Peterson\altaffilmark{7}}

\altaffiltext{1}{Department of Astronomy, School of Science, University
of 
Tokyo, 7-3-1 Hongo, Bunkyo-ku, Tokyo 113-0013, Japan; 
koshida@merope.mtk.nao.ac.jp}
\altaffiltext{2}{National Astronomical Observatory, 2-21-1 Osawa,
Mitaka, 
Tokyo 181-8588, Japan}
\altaffiltext{3}{Institute of Astronomy, School of Science, University
of 
Tokyo, 2-21-1 Osawa, Mitaka, Tokyo 181-0015, Japan}
\altaffiltext{4}{Research Center for the Early Universe, School of
Science, 
University of Tokyo, 7-3-1 Hongo, Bunkyo-ku, Tokyo 113-0013, Japan}
\altaffiltext{5}{Institute of Space and Astronomical Science, Japan
Aerospace 
Exploration Agency, 3-1-1, Yoshinodai, Sagamihara, Kanagawa 229-8510,
Japan}
\altaffiltext{6}{Kiso Observatory, Institute of Astronomy, School of
Science, 
University of Tokyo, 10762-30 Mitake, Kiso, Nagano 397-0101, Japan}
\altaffiltext{7}{Mount Stromlo Observatory, Research School of Astronomy
and 
Astrophysics, Australian National University, Weston Creek P.O., ACT
2611, 
Australia}

\begin{abstract}

The long-term optical and near infrared monitoring observations for a type 1 active galactic nucleus NGC 4151 were carried out for six years from 2001 to 2006 by using the MAGNUM telescope, and delayed response of flux variations in the $K(2.2\mu m)$ band to those in the $V(0.55\mu m)$ band was clearly detected.  Based on cross correlation analysis, we precisely measured a lag time $\Delta t$ for eight separate periods, and we found that $\Delta t$ is not constant changing between 30 and 70 d during the monitoring period.  Since $\Delta t$ is the light travel time from the central energy source out to the surrounding dust torus, this is the first convincing evidence that the inner radius of dust torus did change in an individual AGN.  In order to relate such a change of $\Delta t$ with a change of AGN luminosity $L$, we presented a method of taking an average of the observed $V$-band fluxes that corresponds to the measured value of $\Delta t$, and we found that the time-changing track of NGC 4151 in the $\Delta t$ versus $L$ diagram during the monitoring period deviates from the relation of $\Delta t \propto L^{0.5}$ expected from dust reverberation. This result, combined with the elapsed time from period to period for which $\Delta t$ was measured, indicates that the timescale of dust formation is about one year, which should be taken into account as a new constraint in future studies of dust evolution in AGNs. 

\end{abstract}


\section{Introduction}

The unified model of active galactic nuclei (AGNs) \citep[][]{anto93,urry95} assumes the existence of dust torus that surrounds the central hierarchical structure consisting of a super massive black hole, an accretion disk, and a broad line region (BLR).  The dust grains in the torus absorb the UV/optical continuum emission from the accretion disk and re-radiate in the near infrared (IR) wavelength region with some lag time corresponding to the light travel time from the accretion disk to the inner radius of dust torus.  Since the heated dust eventually sublimates at a constant temperature of 1500-1800K, a more luminous AGN should have a larger dust torus and hence a larger lag time, yielding a correlation of $\Delta t \propto L^{0.5}$ between lag time $\Delta t$ and AGN luminosity $L$ \citep[][]{barv87, barv92}.

In fact, based on the long-term multicolor monitoring data from the MAGNUM project \citep{yosh02,yosh03} and available archival data, \citet{suga06} recently reported such a correlation from a sample of AGNs spanning a wide range of absolute $V$-magnitude from $M_V= -15$ to $-24$.  The optical luminosity is a good indicator of UV luminosity, because their variations are well synchronized with each other \citep{edel96}. Therefore, the reported correlation in the $V$ band indicates that the inner radius of dust torus scales with the UV luminosity as well, which provides a strong piece of evidence for dust reverberation proceeding in the central region of AGNs.

Given the above result of scaling among AGNs of different luminosities, it is possible to examine how $\Delta t$ would scale with changing $L$ in an individual AGN.  This is currently an only diagnosis on its inner structure which is too small to be resolved by direct imaging.  However, since a range of flux variation in an AGN is as small as $<0.1$ mag, detection of a change of $\Delta t$ is indeed a challenge and needs an elaborate method that enables precise estimation of $\Delta t$. 

We here attempt to detect, if any, a change of $\Delta t$ for the case of a nearby Seyfert 1 AGN of NGC 4151, using the six year log accurate and frequent monitoring data for bright AGNs in the program of MAGNUM.  

In Section \ref{sec:app}, we show details of observation of NGC4151. In Section \ref{sec:mod} and Section \ref{sec:tes} we formulate estimation of $\Delta t$ and $L$, by making use of model light curves.  In Section \ref{sec:ana} we apply the formulation to the case of NGC 4151 and derive its time-changing track in the $\Delta t$ versus $L$ diagram. In Section \ref{sec:dis} we discuss time scale of dust formation in the central region of NGC 4151.

\section{Observations of NGC 4151}\label{sec:app}

NGC4151 is one of the bright AGNs most frequently observed since 2001 by the MAGNUM telescope. The monitoring period analyzed in this Letter covers a length of $2000$ d from MJD $52,000$ to $54,000$ (2001-2006), and photometric images were obtained for about 220 nights in the $V$ and $K$ bands. The observations, photometry, and data analysis are mostly in common with those in \citet{mine04}, and we do not repeat them except for some important basics below. 


Each night we took images of the AGN and the reference star alternatively using the multicolor imaging photometer (MIP) with the viewing angle of $1'.5$, and performed relative photometry of the AGN with an aperture size of $\phi=8''.3$.  This size was chosen to minimize the photometric error and the flux variation in the aperture by the seeing effect.

The flux in the aperture contains the flux from the host galaxy, especially from its bulge.  The host component was estimated by the model fitting to its surface brightness as $f_{V,\textrm{gal}}=17.95~\mathrm{mJy}$ \citep{sakapre} and $f_{K,\textrm{gal}}=44.22~\mathrm{mJy}$ \citep{mine04}, and we subtracted these contributions from the $V$ and $K$ light curves, respectively.  The narrow line flux was also significant in the $V$ band for NGC4151, so we subtracted $10.41$ mJy following \citet{sakapre}.  Furthermore, the IR flux contains the flux from the accretion disk, which makes $\Delta t$ systematically shorter than the actual lag.  According to \cite{tomi06}, the IR disk component should vary to synchronize with the optical variation that is originated genuinely from the accretion disk. This component in the $K$ band was then estimated as $f_{K,\textrm{acc}}=(\nu_{V}/\nu_{K})^{\alpha} f_{V}$, and we also subtracted this contribution from the $K$ light curve by assuming $\alpha=0$ for the disk 
emission between the $V$ and $K$ bands. 

The resulting $V$ and $K$ light curves of NGC 4151 from 2001 to 2006 are shown in Figure \ref{fig:fig2}.  The $V$ light curve shown here is consistent with the optical continuum light curves reported by \citet{shap08}. We see that the optical flux of NGC 4151 reached a minimum around MJD $52,000$ and then increased until reaching a maximum at MJD $52,700$.  Thereafter, it turned to decrease again to a minimum at MJD $53,500$. Apparently, the flux variation in the $K$ band followed that in the $V$ band with clear time delay.  

\section{Method of Analysis}\label{sec:mod}

The light curve shown in Figure \ref{fig:fig2} is complex, with many peaks and valleys. From it we need to extract a table of $\Delta t$ as a function of continuum flux which has sufficient precision for us to test the hypothesis that $\Delta t$ is proportional to $f_{V}^{0.5}$ as the flux varies, and/or to understand any difference from that expected result.

The techniques used previously are not adequate. In our series of papers on dust reverberation in AGNs \citep{mine04,suga06,tomi06}, we estimated $\Delta t$ and its error based on the Cross Correlation Function (CCF) analysis by making use of the structure function \citep[SF,][]{whit94} for the interpolation of optical flux variability during the monitoring period. In the equal sampling (ES) scheme of interpolation which we used for the CCF analysis \citep{suga06}, the number of data points generated at equal intervals of time can exceed that actually observed. In such a case, the simulated light curve shows larger flux variation compared to an overall trend seen from the observed light curve, especially when the data points are generated far from the location of observed epochs. This necessarily gives rise to too large an error of $\Delta t$.

We here adopt the bi-directional interpolation (BI) scheme, where all $V$-band flux data are paired with $K$-band flux data and at least either $V$ or $K$ flux in each pair is measured by actual observation.  While the BI scheme overcomes a problem of generating too many data points for CCF calculations, the CCF still shows an unreal peak with increasing the number of data pairs. This drawback is remedied by estimating $\Delta t$ from the CCF centroid around the peak which is interpreted as a luminosity-weighted distance of the dust torus from the AGN central energy source \citep{kora91}. As a practice of determining the CCF range for centroid estimation, the threshold is usually set to be 0.8 times the CCF-peak value. If the threshold is larger than $0.9$, $\Delta t$ is influenced by an unreal peak of the CCF. On the other hand, if it is smaller than $0.7$, $\Delta t$ is necessarily influenced by sub-features such as a flared foot of the CCF. 

We so far used a simple mean of observed $V$-band fluxes as an indicator of optical luminosity that corresponds to $\Delta t$, but this could seriously be biased by the two factors.  First, the flux mean depends on the sampling frequency of light curve during the monitoring period.  When an AGN is often observed in the bright phase rather than in the faint phase, the simple flux mean preferentially gives a value of brighter flux.  Second, the flux mean depends on the shape of light curve.  When an AGN undergoes an abrupt burst on a flat light curve, the simple flux mean gives a value which is more or less the same as the value of flat part of the light curve, although the flat part contributes very little to the value of cross-correlation (CC) coefficient for the estimation of $\Delta t$.

Considering above, we take a mean of $V$-band fluxes generated at equal intervals of time, after excluding the data which statistically show no flux variation in time bins.  To pick up such data, we use the equally interpolated optical light curve for which the size of time bins is taken as the mean interval of monitoring observations.  The variability in each bin is tested by $\chi^2$ with a significance level of $0.99$, and the data which are judged as showing no flux variation are excluded in calculating the mean of $V$-band fluxes.  The distribution of this flux mean for all simulated light curves is then used to derive an average optical flux $f_V$ and its error.

\section{A Test Using Model Light Curves}\label{sec:tes}

We next use model light curves to verify that our analysis procedure reliably finds a characteristic lag $\Delta t$ and its associated flux $f_{V}$ for each of the individual variability events that combine to make the complex light curve shown in Figure 1.  In order to quantify the above estimation of a characteristic lag $\Delta t$ and its associated flux $f_V$ as well as their errors, we model the light curves to calculate CCFs. Given a $V$ light curve $f_V(t)$, we construct a corresponding $K$ light curve as $f_K(t)=f_V(t-\Delta t)$, where $\Delta t/50=(f_V/200)^{\gamma}$ for $\gamma =0.5\pm0.3$ which roughly reproduce the lags seen in Figure \ref{fig:fig2}.  Here, the units are days in time and arbitrary in flux.  Then we generate 30 data points along each of the $V$ and $K$ light curves, and disperse the points according to a Gaussian distribution of photometric errors taken from the actual observations by the MAGNUM telescope.    

The models include three characteristic shapes of (1) linear bottom, (2) quadratic bottom, (3) bursting Gaussian peak, as shown in order from top to bottom in the left column of Figure \ref{fig:fig1}.  The models 1 to 3 are shown in Figure \ref{fig:fig1} to examine the dependence on shape of time-symmetric incident $V$ light curve used for CCF calculation.  In addition to all these models with a standard index of $\gamma=0.5$, we consider two cases of $\gamma=0.2$ and $0.8$ for each of light curve shapes 1 and 2. We note, however, that $\gamma>0.5$ is considered only for academic purpose and contradicts with the geometrical $1/r^2$-dilution of the incident flux from the central energy source. 

Following \citet{suga06}, we applied the BI method to each of model light curves to calculate the CCF and also the CC centroid distribution (histogram) by Monte-Carlo simulations, as shown in the middle column of Figure \ref{fig:fig1} for $\gamma =0.5$.  The results of $\Delta t$ and its error are given on the panels of this column and in Table \ref{tab:tab1}. 
The results of $\Delta t$ for the models 1 to 3 give values around $80$ d. This indicates that $\Delta t$ could be estimated independently of the shape of light curve.  

Now we derive an average flux $f_V$ for each of the models 1 to 3, following the way described in Section \ref{sec:mod}.  The horizontal arrows in the left column of Figure \ref{fig:fig1} specify the periods in which the data are rejected in the estimation of flux mean due to their invariability.  We calculate the distribution of this flux mean for the light curves generated by Monte-Carlo simulations, and present it as a histogram from which $f_V$ and its error are derived, as shown in the right column of Figure \ref{fig:fig1} for $\gamma =0.5$. The results of $f_V$ and its error are given on the panels of this column and in Table \ref{tab:tab1}.

To be consistent with the measured value of $\Delta t$, we compare our estimate of $f_V$ with that from the relation $\Delta t/50=(f_V/200)^{\gamma}$ adopted in the CCF analysis.  The relative difference of $\delta f_V/f_V \equiv (f_V -f_V(\Delta t))/f_V(\Delta t)$ is shown in Table \ref{tab:tab1}.  We see that $|\delta f_V/f_V|$ for $\gamma=0.5$ is less than $10\%$ for all the models 1 to 3, which is considerably improved over our previous estimate of $40\%$ by using the simple flux mean.   

The result of $f_V$ for different values of $\gamma=0.2$ and $0.8$ tabulated in Table \ref{tab:tab1} shows little difference from that for $\gamma=0.5$, which indicates that estimated $f_V$ is almost independent of this parameter.  In addition, various other light curves in asymmetric shapes also give an essentially same result.

\section{Application to NGC4151}\label{sec:ana}

Given that the consistent estimation of $\Delta t$ and $f_V$ has been confirmed as above by using the model light curves, we apply it to the real data for NGC 4151. We divided the whole monitoring period into eight separate periods, each of which contains a single feature of flux variation, as above, with a sufficient number of more than 40 flux data pairs for the CCF analysis.  The middle date of each period is taken as an epoch that represents each period.  Following the procedure described in Section \ref{sec:mod}, the Monte-Carlo simulation was repeated 500 times for each period to form the distribution of CC centroid lag from which $\Delta t$ was obtained as the median and its error as $\pm$ 34.1 percentile from the median.  Similarly, we calculated the distribution of flux mean from which $f_V$ and its error were obtained. Their results for eight separate periods are given in Table \ref{tab:tab2} as well as in Figure \ref{fig:fig3}.

We here note that the $K$-band data of NGC 4151 were well sampled and the observed $K$ light curves were smooth enough.  Therefore, in order to avoid additional IR flux variation which consequently overestimates the errors in $\Delta t$, use of the SF for the interpolation was turned off and the simple linear interpolation was adopted for the $K$-band data in making pairs with the $V$-band data. 

It is apparent that $\Delta t$ is not constant in time.  For the most extreme case, the difference of $\Delta t$ between the earlier periods of 1 through 4 and the later periods of 6 through 8 is larger than the measurement error by many $\sigma$s.  Since $\Delta t$ corresponds to a light-travel distance to the surrounding dust from the AGN center, this should be the first convincing evidence that the inner radius of dust torus did change  in an individual AGN \citep[cf.][]{okny99,mine04}.

\section{Discussion}\label{sec:dis}

The change of inner radius of dust torus should occur reflecting the variation of the incident UV/optical flux from the central energy source.  Some theoretical models suggest the existence of sublimation radius inside which dust grains are sublimated \citep{barv87,barv92}.  The sublimation radius expands when the UV/optical continuum becomes bright, then it retreats when the continuum becomes faint.  If dust grains were sublimated or replenished immediately after the UV/optical flux variation, like the model we assumed in Section \ref{sec:mod}, the expected changes of $\Delta t$ and $f_V$ should trace the simple relation of $\Delta t \propto L^{0.5}$ which has the slope indicated by the thick line in Figure \ref{fig:fig3}.  We see from this figure that the observed change of $\Delta t$ follows the variation of $f_V$ in overall trend as expected.  These results indicate that the dust torus is not a distinct, separate physical structure but is a part of continuous structure starting from the BLR component out to dust component with its inner dust edge set by the sublimation radius that changes with the UV/optical flux variation \citep{suga06, nenk08}. 

On the other hand, however, the observed time-changing track does not exactly trace the simple relation of $\Delta t \propto L^{0.5}$.  The lag $\Delta t$ at the period 1 is well above that expected for $M_V$, and does not change so much with brightening of about 1 mag from the period 1 to 3. Then, with subsequent fading from the period 3 to 6, $\Delta t$ still remains constant until the period 4, and starts to rapidly decrease thereafter.  These deviant behaviors suggest that the inner dust torus did not reach an equilibrium state immediately after the UV/optical flux variation.  

\begin{figure}
\includegraphics[scale=0.8]{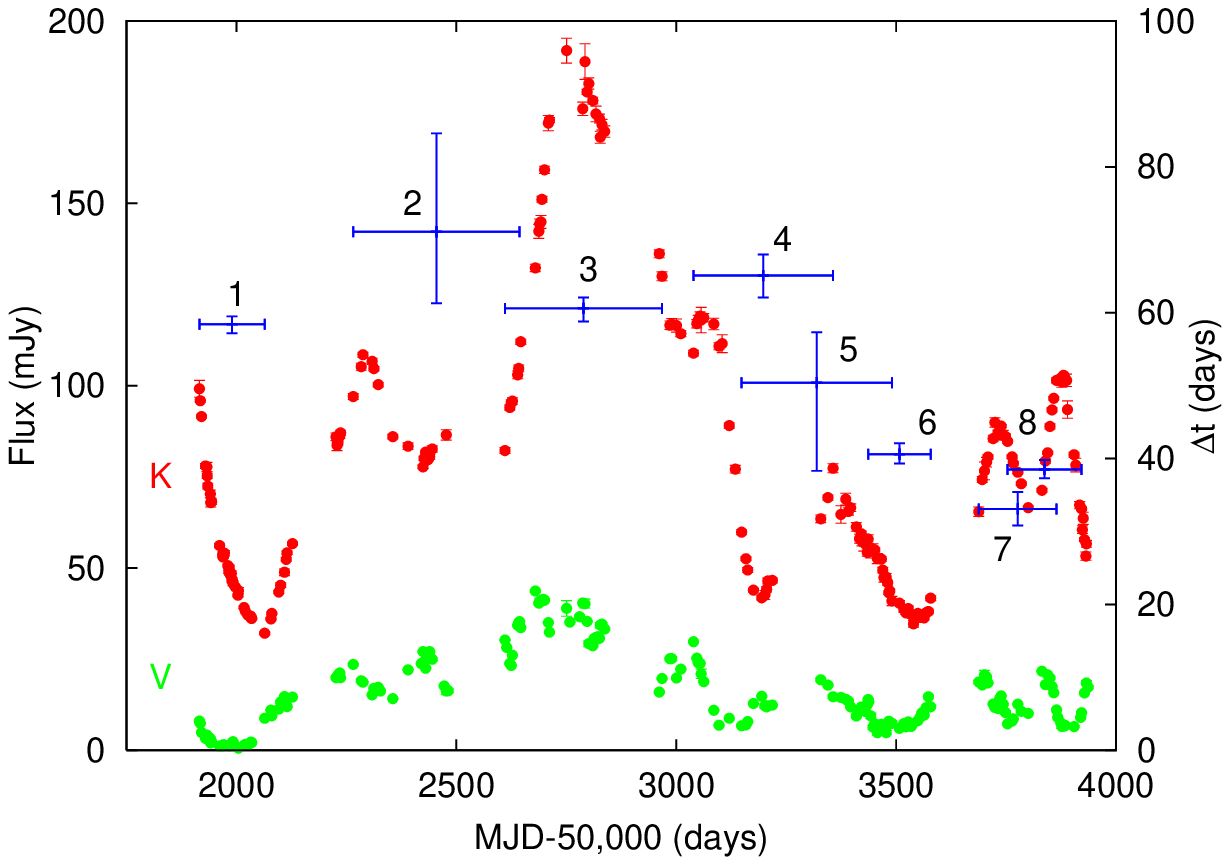}
\caption{The $V$ and $K$ light curves of NGC 4151 (dots) in units of mJy along the vertical axis on the left, and the lag times $\Delta t$ (crosses) in units of days along the vertical axis on the right. The horizontal bar to each cross does not specify the error but the time interval for which $\Delta t$ is estimated, and the vertical bar indicates the error of $\Delta t$ by the CCF simulations.  The number beside each cross refers to the serial number given to each monitoring period in Table \ref{tab:tab2}.}
\label{fig:fig2}
\end{figure}

\begin{figure}
\includegraphics[scale=0.8]{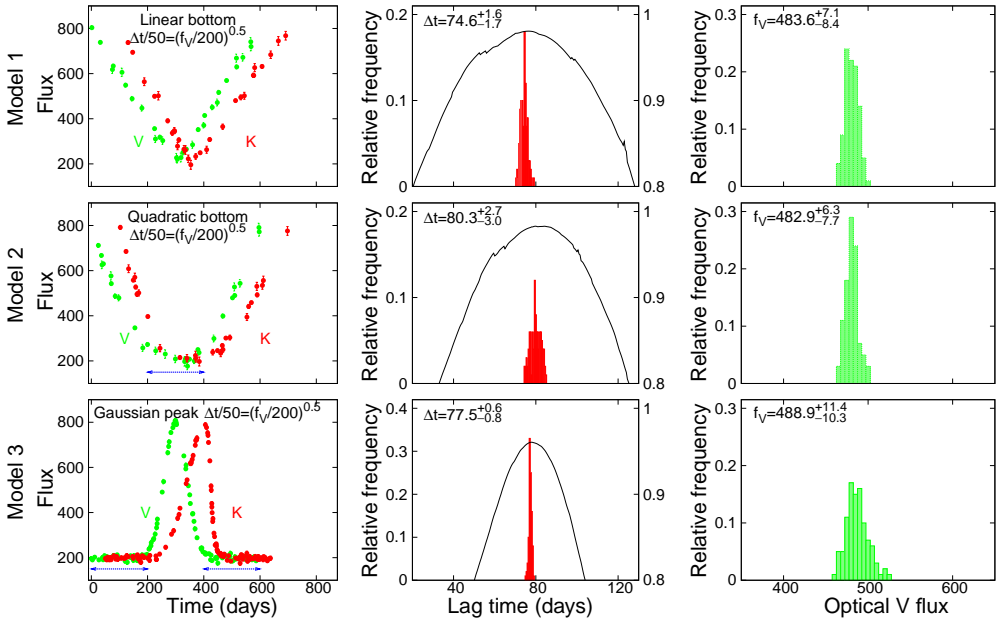}
\caption{Models of $V$ and $K$ light curves (left column), and the corresponding distributions of CC centroid lag (middle column) and mean $V$ flux (right column), for the cases of time-symmetric incident $V$ light curve.  Three panels from left to right in each row shows a procedure of estimating $\Delta t$ and $f_V$ from each model of light curve (see text).  The solid lines on the panels of middle column show the CCFs with their values on the right axis.  The horizontal arrows on the panels of left column indicate the periods for which the flux data are excluded for estimation of $f_V$.  Note that the units are arbitrary in flux.}
\label{fig:fig1}
\end{figure}

\begin{figure}
\includegraphics[scale=0.8]{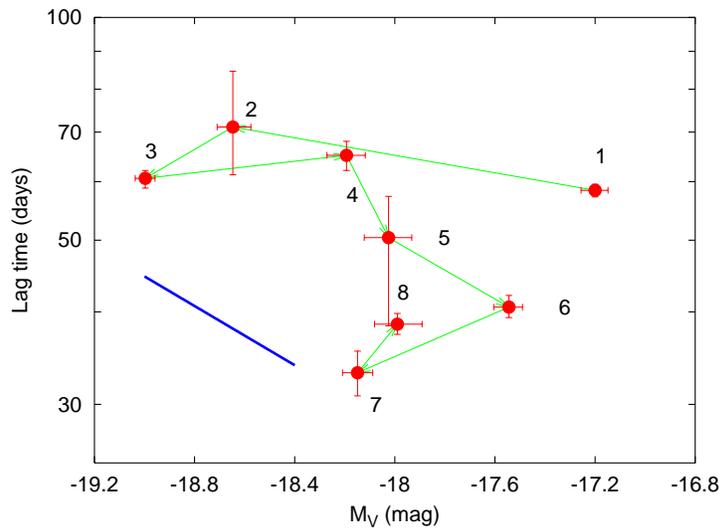}
\caption{The time-changing track of NGC 4151 in the lag time versus absolute $V$-magnitude diagram, as shown by the dot-connecting thin line. The number beside each dot refers to the serial number given to each monitoring period in Table 2. The thick line is the scaling relation of $\Delta t \propto L^{0.5}$ expected from dust reverberation. We used $H_0=70$km~s$^{-1}$~Mpc$^{-1}$ for estimation of $M_V$.} 
\label{fig:fig3}
\end{figure}

In the case of period 1, the light curves in Figure \ref{fig:fig2} indicate that there was a brighter period before the period 1.  The expanded inner radius of dust torus at the brighter period could still remain larger than the radius that would be expected at the period 1.  In this context, the behavior of almost constant $\Delta t$ followed by its rapid decrease in the fading period places a constraint such that a time scale of dust replenishment in the central region of AGN should be as long as the elapsed time of about one year from the period 4 through 6.  A promising mechanism of such replenishment is re-formation of dust grains rather than their infall from the outer region.  This is understood by considering that the decrease of $\Delta t$ from the period 4 to 6 is 24.5 d or a light-travel distance of $6.35\times 10^{11}$ km in the interval of 309 d between the middle dates of these periods.  If such decrease of $\Delta t$ would occur with redistribution of dust grains supplied from the outer region of dust torus, the infall velocity would be $2.4\times 10^{4}$ km~s$^{-1}$ or 7.9\% of the light velocity.  This seems highly unlikely, when we compare this infall velocity with a few $\times 10^{3}$ km~s$^{-1}$ for the velocity dispersion of BLR clouds which exist just near the inner dust torus.  Consequently, we conclude that re-formation of dust grains did occur in the central region where they had been sublimated.

After the dust re-formation period, $\Delta t$ turned almost constant again from the period 6 through 8. This indicates that dust sublimation radius became larger than actual inner radius of dust torus and dust sublimation would start.  The length of period of constant $\Delta t$ suggests that a time scale of dust destruction could be longer than about one year.  Exact measurement of this time scale might have been possible if the light curves were obtained well after the period 8.

The formulation for improving the accuracy of lag and luminosity measurements was applied to the light curves for eight separate periods containing single features.  This formulation, when applied to the light curves for the whole monitoring period containing several features, gives an almost average of their respective $\Delta t$ estimates derived from single features, and is still useful in discussing an overall lag--luminosity relationship for a sample of many AGNs with a wide range of luminosity.

\acknowledgments

We thank the staff at the Haleakala Observatories for their help with facility maintenance.  This research has been supported partly by the Grants-in-Aid of Scientific Research (10041110, 10304014, 11740120, 12640233, 14047206, 14253001, 14540223, and 16740106) and the COE Research (07CE2002) of the Ministry of Education, Science, Culture and Sports of Japan.

\clearpage

{\renewcommand \arraystretch{1.2}
\begin{deluxetable}{cllllr}
\tabletypesize{\scriptsize}
\tablecaption{Results of simulation runs for model light curves}
\tablewidth{0pt}
\tablehead{
\colhead{Model} & \colhead{Type} & \colhead{$\gamma$} & \colhead{$\Delta t$} 
(days) &
 \colhead{$f_V$} & 
\colhead{$\delta f_V/f_V$}
}
\startdata
1 & Linear bottom & 0.5 & $74.6{}^{+1.6}_{-1.7}$ &
$483.6{}^{+7.1}_{-8.4}$ & 
$0.09$ \\
2 & Quadratic bottom & 0.5 & $80.3{}^{+2.7}_{-3.0}$ &
$482.9{}^{+6.3}_{-7.6}$ 
& $-0.06$ \\
3 & Gaussian peak & 0.5 & $77.5{}^{+0.6}_{-0.8}$ &
$488.9{}^{+11.4}_{-10.3}$ & 
$0.02$ \\ \tableline
1a & Linear bottom & 0.2 & $59.1{}^{+2.7}_{-2.3}$ &
$489.4{}^{+9.2}_{-7.3}$ & 
$0.06$ \\
2a & Quadratic bottom & 0.2 & $59.9{}^{+1.9}_{-1.8}$ &
$475.2{}^{+9.7}_{-5.5}$ & 
$-0.04$ \\ \tableline
1b & Linear bottom & 0.8 & $107.3{}^{+2.3}_{-2.3}$ &
$490.9{}^{+7.6}_{-7.7}$ & 
$-0.06$ \\
2b & Quadratic bottom & 0.8 & $107.1{}^{+5.2}_{-4.3}$ &
$481.6{}^{+7.8}_{-9.7}$ & 
$-0.05$
\enddata
\tablecomments{
$\gamma$ is a power index for the generator $\Delta 
t/50=(f_{V}/200)^{\gamma}$.  $\delta f_V/f_V$ is defined as 
$(f_V -f_V(\Delta t))/f_V(\Delta t)$, where $f_V(\Delta t)$ is 
estimated from the generator with the measured value of $\Delta t$. 
The units here are days for $\Delta t$ and arbitrary for $f_V$. 
}
\label{tab:tab1}
\end{deluxetable}
}

{\renewcommand \arraystretch{1.2}
\begin{deluxetable}{clll}
\tabletypesize{\scriptsize}
\tablecaption{Lag time and $V$-magnitude of NGC 4151 
for separate monitoring periods}
\tablewidth{0pt}
\tablehead{
\colhead{Period} & \colhead{MJD (days)} & \colhead{$\Delta t$ (days)} & \colhead
{$m_{V}$ (mag)}
}
\startdata
1 & $51915.6$--$52064.3$ & $58.4_{~~-1.2}^{~~+1.1}$ & $14.49~~_{-0.06}^{+0.05}$ 
\\
2 & $52265.5$--$52643.6$ & $71.1_{~~-9.8}^{~+13.5}$ & $13.04~~_{-0.07}^{+0.07}$ 
\\
3 & $52610.6$--$52967.6$ & $60.6_{~~-1.8}^{~~+1.5}$ & $12.69~~_{-0.04}^{+0.04}$ 
\\
4 & $53039.4$--$53356.5$ & $65.1_{~~-3.0}^{~~+2.9}$ & $13.50~~_{-0.08}^{+0.07}$ 
\\
5 & $53148.4$--$53490.3$ & $50.4_{~-12.1}^{~~+6.9}$ & $13.66~~_{-0.10}^{+0.09}$ 
\\
6 & $53436.4$--$53578.3$ & $40.6_{~~-1.3}^{~~+1.5}$ & $14.15~~_{-0.06}^{+0.05}$ 
\\
7 & $53687.6$--$53864.4$ & $33.1_{~~-2.3}^{~~+2.3}$ & $13.54~~_{-0.06}^{+0.06}$ 
\\
8 & $53753.4$--$53921.3$ & $38.5_{~~-1.2}^{~~+1.3}$ & $13.70~~_{-0.10}^{+0.10}$
\enddata
\label{tab:tab2}
\end{deluxetable}
}


\begin{thebibliography}{}
\bibitem[Antonucci(1993)]{anto93} Antonucci, R., 1993, \araa, 31,473
\bibitem[Barvainis(1987)]{barv87} Barvainis, R., 1987, \apj, 320, 537
\bibitem[Barvainis(1992)]{barv92} Barvainis, R., 1992, \apj, 400, 502
\bibitem[Edelson et al.(1996)]{edel96} Edelson, R. A. et al., 1996, \apj, 470, 364
\bibitem[Koratkar \& Gaskell(1991)]{kora91} Koratokar, A.~P., \& Gaskell, C.~M., 1991, \apjs, 75, 719
\bibitem[Minezaki et al.(2004)]{mine04} Minezaki, T., Yoshii, Y., Kobayashi, Y., Enya, Keigo., Suganuma, M., Tomita, H.,  Aoki, T., \& Peterson, B. A., 2004, \apj, 600, L35
\bibitem[Nenkova et al.(2008)]{nenk08} Nenkova, M., Sirocky, M. M., Nikutta, R., Ivesi\'{c}, \v{Z}., \& Elitzur, M., 2008, \apj, 685, 160
\bibitem[Oknyanskij et al.(1999)]{okny99} Oknyanskij et al. 1999, Astronomy Letters, 25, 483
\bibitem[Shapovalova et al.(2008)]{shap08} Shapovalova, A. I. et al. 2008, \aap, 486, 99
\bibitem[Suganuma et al.(2006)]{suga06} Suganuma, M. et al., 2006, \apj, 639, 46
\bibitem[Tomita et al.(2006)]{tomi06} Tomita, H. et al. 2006, \apjl, 652, L13
\bibitem[Urry and Padovali(1995)]{urry95} Urry, C. \& Padovani, P. 1995, \pasp, 107, 803
\bibitem[White \& Peterson(1994)]{whit94} White, R. J., \& Peterson, B. M., 1994, \pasp, 106, 879
\bibitem[Yoshii(2002)]{yosh02} Yoshii, Y., 2002, in New Trends in Theoretical and Observational Cosmology, ed. K. Sato \& T. Shiromizu (Tokyo: Universal Academy Press), 235
\bibitem[Yoshii et al.(2003)]{yosh03} Yoshii, Y., Kobayashi, Y., \& Minezaki, T., 2003, \baas, 202, 38.03
\end{thebibliography}
\end{document}